\documentclass[aps,prb,twocolumn,showpacs,floats]{revtex4-1}
\usepackage[dvips]{graphicx}
\usepackage{amsmath,amssymb,amsfonts,bm,graphicx,hyperref}
\usepackage{empheq}
\usepackage{dsfont}

\newcommand{\ra}{\rangle}
\newcommand{\la}{\langle}
\newcommand{\ud}{\mathrm{d}}
\begin{document}

\title{\bf A quantum magnetic RC circuit}

\author{Kevin A. van Hoogdalem$^1$, Mathias Albert$^2$, Pascal Simon$^2$, and Daniel Loss$^1$}

\affiliation{$^1$Department of Physics, University of Basel, Klingelbergstrasse 82, CH-4056 Basel, Switzerland}
\affiliation{$^2$Laboratoire de Physique des Solides, CNRS, Universit\'e Paris-Sud, 91405 Orsay, France}

\date{\today}

\begin{abstract}
We propose a setup that is the spin analog of the charge-based quantum RC circuit. We define and compute the spin capacitance and the spin resistance of the circuit for both ferromagnetic (FM) and antiferromagnetic (AF) systems. We find that the antiferromagnetic setup has universal properties, but the ferromagnetic setup does not. We discuss how to use the proposed setup as a quantum source of spin excitations, and put forward a possible experimental realization using ultracold atoms in optical lattices.
\end{abstract}

\maketitle
{\it Introduction}.
The aim in the field of spintronics in insulating magnets\cite{Meier2003, Kostylev2005, Imre2006, Kajiwara2010, Uchida2010, Khajetoorians2011, Menzel2012, Hoogdalem2013,Tserkovnyak2013} is to use purely magnetic collective excitations, such as magnons\cite{Mattis1981} or spinons~\cite{Giamarchi2003}, to perform logic operations in the absence of charge transport. Thereby, it is possible to circumvent the problem of excess Joule heating that occurs due to the scattering of conduction electrons in more traditional electronic devices, leading to lower energy dissipation in such spintronic devices\cite{Trauzettel2008}. Since this excess heating is a limiting factor in the design of electronic devices, spintronics in insulating magnets is considered one of the candidates to become the next computing paradigm. Furthermore, the fact that the elementary excitations in ferromagnetic insulators obey bosonic statistics may offer additional benefits\cite{Meier2003}.

Several experiments that display the capability to create and detect pure spin currents in magnetic insulators  have been performed recently. Creation of a magnon current has been shown to be possible using the spin Hall effect~\cite{Kajiwara2010}, the spin Seebeck effect~\cite{Uchida2010}, as well as laser-controlled local temperature gradients~\cite{Weiler2012}; detection of magnon currents has been performed using the inverse spin Hall effect~\cite{Kajiwara2010,Sandweg2011}. However, analogously to 
quantum optics where the single-photon source is a major element to encode or manipulate a quantum state~\cite{Gisin2002}, or to quantum electronics where an on-demand electron source has been recently realized~\cite{Feve2007,Parmentier2010,Dubois2013}, a more controllable way of creating quantum spin excitations may ultimately be desirable.

Besides offering great potential for applications, single-excitation sources are also fascinating from a more fundamental point of view. This is illustrated by the single-electron source, which violates the classical laws of electricity\cite{Buttiker1993,Gabelli2006}. Furthermore, in the linear response regime and at low driving-frequency, a single-electron source can be described in terms of a quantum RC circuit whose charge relaxation resistance has a universal value \cite{Buttiker1993,Nigg2006,Nigg2008,Mora2010,Hamamoto2010,Filippone2011,Filippone2012,Gabelli2012}.
Motivated by these considerations, we analyze here a setup that we propose could potentially act as an on-demand coherent source of magnons or spinons and compute the equivalent RC parameters of such a circuit from a microscopic model. 

\begin{figure}[h!]
\centering
\includegraphics[width=0.95\columnwidth]{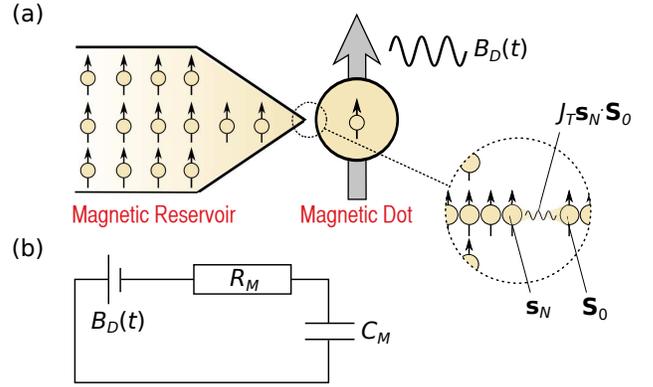}
\caption[]{(a) Schematic representation of the setup. The weakly coupled reservoir and dot are both modelled as 1D chains in this work. Parallel collections of such 1D chains are realized in bulk materials such as SrCuO$_2$ (Ref. \onlinecite{Hlubek2010}) and Cs$_2$CoCl$_4$ (Ref. \onlinecite{Kenzelmann2002}), as well as in ultracold atoms in optical lattices\cite{Kuklov2003,Duan2003}. (b) Equivalent circuit representation of the setup, see Eq. (\ref{eq:response}) for the definition of $R_M$ and $C_M$.}
\label{fig:Fig1}
\end{figure}

By drawing analogy to the charge-based quantum RC circuit, we propose that the setup depicted in Fig. \ref{fig:Fig1}(a) is equivalent to a `quantum magnetic RC circuit'. We mean by this that in the displayed setup $M_D(\omega)$, the excess magnetization of the magnetic grain or magnetic dot (see below), is related to the applied magnetic field $B_D(\omega)$ by
\begin{equation}
\frac{M_D(\omega)}{B_D(\omega)} = C_M \left(1 + i \omega C_M R_M\right).
\label{eq:response}
\end{equation}
Here $C_M$ and $R_M$ are the magnetic resistance and capacitance of the equivalent RC circuit [see Fig. \ref{fig:Fig1}(b)]. We emphasize that our proposed magnon or spinon source is not equivalent to a classical spin battery \cite{Brataas2002} but operates at the quantum level (\textit{i.e.} on the level of individual coherent magnons/spinons).

The excess magnetization of the nonitinerant magnetic dot  is defined as $M_D(\omega) = g\mu_B N_D(\omega)$, where $N_D(\omega)$ is the excess number of magnetic quasiparticles in the dot. These quasiparticles are the elementary quantum excitations (magnons or spinons) of the Heisenberg Hamiltonian which we will use to describe the dot. We must make a clear distinction between ferromagnetic- (FM) and antiferromagnetic (AF) systems here. The main difference between the two lies in the different statistics obeyed by the respective elementary excitations; whereas the FM magnons obey bosonic statistics, the AF spinons in contrast behave according to fermionic statistics. This leads us to expect very different behaviour between these systems.

{\it Model}.
Our setup consists of a magnetic dot or magnetic grain that is weakly exchange-coupled to a large magnetic reservoir. Both the magnetic dot and  reservoir are assumed to be nonitinerant magnets, described by a Heisenberg Hamiltonian. For concreteness, we will model our subsystems as 1D spin chains. We characterize the system by the Hamiltonian $H = H_0 + H_{T}$. Here, $H_0 = H_D+ H_R$ describes the isolated subsystems and $H_{T}$ the weak magnetic exchange interaction between the dot and reservoir. The Heisenberg Hamiltonian $H_D$ that describes the isolated magnetic dot is given by
\begin{equation}
H_D = \sum_{\la i j\ra} {\bf S}_i \cdot {\bf J}_D \cdot {\bf S}_j +  g\mu_B \sum_i \left[{\bf B}_D^0 + {\bf B}_D(t) \right]\cdot {\bf S}_i.
\label{eq:Ham}
\end{equation}
${\bf J}_D$ denotes a diagonal $3\times 3$-matrix with $\textrm{diag}({\bf J}_D) = J_D\{ 1, 1, \Delta_D\}$. $J_D$ is the magnitude of the exchange interaction and $\Delta_D$ the anisotropy in our model. $J_D \lessgtr 0$ corresponds respectively to the FM and the AF ground state. We note that anisotropy is typically caused by spin-orbit- or dipole-dipole interactions. ${\bf B}_D^0 = B_D^0 {\bf e}_z$ and ${\bf B}_D(t) = B_D(t) {\bf e}_z$ are respectively the static- and time-dependent component of the magnetic field applied to the dot. The Hamiltonian $H_R$ that describes the reservoir is given by Eq. (\ref{eq:Ham}) with parameters ${\bf J}_R, J_R, \Delta_R$, $B_R^0$, and $B_R(t)=0$. We will use a lowercase ${\bf s}_i$ to denote the $i$th spin in the reservoir. $H_{T}$ will be defined later in Eq. \eqref{eq:Hexch}.

We can either use the Holstein-Primakoff\cite{Mattis1981} (for FM systems) or the Jordan-Wigner\cite{Giamarchi2003} (for spin 1/2 AF systems) transformation to map the spin-ladder operators on respectively bosonic or fermionic creation-/annihilation operators, corresponding to spinless quasiparticles with magnetic moment $g\mu_B {\bf e}_z$ (see appendix). The $z$ component of the spin is mapped onto the density of the respective quasiparticles. Regardless of the statistics of the quasiparticles, we will denote an annihilation operator in the reservoir/dot by $r_i/d_i$. At points where it becomes necessary to distinguish between bosons and fermions, we will do so explicitly.

In thermal equilibrium, the ground state of an isolated dot contains a fixed number $N_0 = \sum_i  \la d^\dagger_i d_i \ra_0$ of magnetic quasiparticles. Here, $\la...\ra_0$ denotes the average with respect to the Hamiltonian $H_0$ with $B_D(t) = 0$. We now define the excess number of magnetic excitations on the dot as $\hat{N}_D = \sum_i d^\dagger_i d_i -N_0$.

\begin{figure}
\centering
\includegraphics[width=0.95\columnwidth]{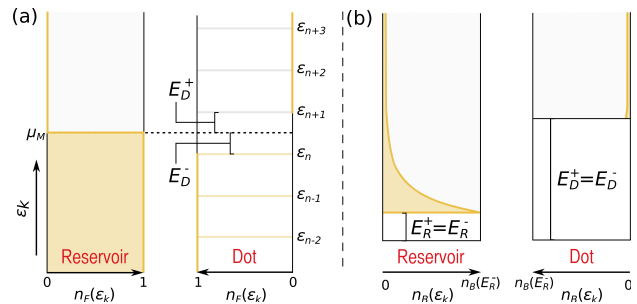}
\caption[]{(a) Band structure of the AF system, assuming a small AF dot with discrete energy levels. Since the magnetic quasiparticles obey fermionic statistics, all states above (below) $\mu_M$ are empty (filled) at $T=0$. We define $E_D^{\pm}=\varepsilon_{n+1/n} - \mu_M$. (b) Band structure of FM system at finite $T \ll E_D^+$. The quasiparticles obey bosonic statistics. Here, $E_{D(R)}^+ = E_{D(R)}^-$ denotes the energy of the lowest energy level in the dot (reservoir).}
\label{fig:Fig2}
\end{figure}

We will consider magnetic dots whose Hamiltonian can be diagonalized as $H_D = \sum_k (\varepsilon_k-\mu_M) d_k^\dagger d_k$, with $\varepsilon_k$ the dispersive energy of the excitations and $\mu_M$ the magnetic equivalent of the chemical potential. The parameters $R_M,C_M$ of the quantum RC model are only well-defined if adding and removing a quasiparticle from the dot involves a finite amount of energy, \textit{i.e.} if the spectrum has a gap (see Fig. \ref{fig:Fig2}). In small magnetic dots of size $L$, quantization of the wave vector $k$ in multiples of $2\pi /L$ leads to a level splitting (and hence $E_D^{\pm}$, see Fig. \ref{fig:Fig2}) of order $J_D(a/L)^2$ for FM dots, and $J_Da/L$ for AF dots ($a$ is the nearest-neighbor distance). 

In principle, an anisotropy $\Delta_D > 1$ gives rise to a bulk gap in large AF dots as well. However, the resulting system is the magnetic equivalent of a Mott insulator, rather than the equivalent of a band insulator\cite{Giamarchi2003}. As a consequence, the excitations are no longer the $d^{(\dagger)}_k$'s of the original model, and the resulting model does not allow for a straightforward analysis. The opening of a bulk gap $E_D^\pm$ by an applied magnetic field requires a staggered field, with wave vector $2k_F \approx \pi/a$. Since neither mechanism allows to create a bulk gap $E_D^{\pm}$ for large AF dots in a straightforward manner, we will rather focus on small AF dots with a finite level splitting due to the quantization of the wave vector $k$ for AF systems. 
The magnetic chemical potential is given by $\mu_M = g\mu_B |B_D^0|$.

For FM dots, we put $\mu_M=0$. There exist two mechanisms that allow for a finite gap even in large FM dots. First, an anisotropy $\Delta_D > 1$ gives rise to a gap $E_D^+ = 2|J_D|S_D(\Delta_D-1)$ (see Fig. \ref{fig:Fig2}(b) for the definition of $E_D^+$ for FM systems). This is due to the fact that the $\Delta_DS_i^zS_j^z$-term maps on a density-density interaction after the transformation to quasiparticles. Second, application of a magnetic field $B_D^0$ leads to $E_D^+ = g\mu_B|B_D^0|$.

For our calculations for FM subsystems, we will assume that the reservoir is described by the isotropic Heisenberg Hamiltonian, {\it i.e.} with $\Delta_R=1$. For AF subsystems, we will assume initially that the reservoir as well as the dot are easy-axis AF spin-1/2 spin chains, {\it i.e.} with $\Delta_{R(D)}=0$. This has the advantage that the excitations can be mapped on free fermions. We will show later how to extend our results for spin chains with finite anisotropy.

The exchange interaction between dot and reservoir is given by $J_{T} {\bf s}_{N}\cdot {\bf S}_{0}$, where ${\bf s}_N$ denotes the last spin in the reservoir, and ${\bf S}_0$ the first spin in the dot. $J_{T}$ is the smallest parameter in the problem, and we analyze the effect of $H_{T}$ using perturbation theory. We show in the appendix that the out-of-plane component of the interaction does not significantly affect our results, so that we can ignore it. This means we can approximate $H_{T}$ by only taking the terms that transfer a magnetic excitation between dot and reservoir into account
\begin{equation}\label{eq:Hexch}
{\hat H}_{T} = J_{T}[ r_N^\dagger d_0 + r_Nd^\dagger_0].
\end{equation}

By using linear-response theory, we can calculate the change in magnetization $M_D(\omega)$ due to a small time-dependent change in $B_D(\omega)$. It is given by
\begin{equation}
M_D(\omega) = (g\mu_B)^2G_\textrm{ret}(\omega) B_D(\omega).
\end{equation}
The retarded Green's function is given by $G_\textrm{ret}(\omega) = G(\omega) + G^*(-\omega^*)$, where 
\begin{equation}
G(\omega) = i \int_0^\infty \ud t e^{i\omega t} \la \hat{N}_D(t)\hat{N}_D(0)\ra_H
\label{eq:G}
\end{equation}
As usual, $\omega$ contains an infinitesimal imaginary part to ensure convergence of the integral. We will calculate $G(\omega)$ using $\hat{H}_{T}$ as perturbation. A substantial part of the calculations for the FM and the AF setup are identical, and we will distinguish between the two only when necessary.

The correlation function in Eq. (\ref{eq:G}) is given by
\begin{equation}
\la\hat{N}_D(t)\hat{N}_D(0)\ra_H = \frac{\la T_t \hat{U}(-\infty,\infty) \hat{N}'_D(t) \hat{N}'_D(0)\ra_0}{\la  T_t \hat{U}(-\infty,\infty)\ra_0},
\end{equation}
where the prime denotes an operator in the Heisenberg representation with respect to $H_0$ and $\hat{U}(-\infty,\infty) = T_{t_1} \textrm{exp}[-i\int_{-\infty}^\infty \ud t_1 \hat{H}_T'(t_1)]$ is the propagator.
The lowest-order contribution to $G(\omega)$ is quadratic in ${\hat H}_{T}$. It can be written as
\begin{multline}
G(\omega) = -\frac{i}{2} \int_0^\infty \ud t e^{i\omega t} \int_{-\infty}^\infty \ud t_1 \int_{-\infty}^\infty \ud t_2 \\
\times  \la T_t \hat{H}'_{T}(t_1)\hat{H}'_{T}(t_2) \hat{N}'_D(t)\hat{N}'_D(0)\ra_0,
\label{eq:Gsec_order}
\end{multline}
where ${\hat H}_{T}$ should be written in terms of $d_i^{(\dagger)}$'s and $r_i^{(\dagger)}$'s. Since the operator $\hat{N}_D(t)$ is defined such that $\hat{N}'_D(t) | \textrm{gs}\ra = 0$, where $|\textrm{gs}\ra$ denotes the ground state of the system under $H_0$, it follows immediately that there is only one time ordering that gives nonzero contributions (note that $t_1$ and $t_2$ are dummy variables whose exchange does not yield new terms). This time ordering leads to two different contributions to $G(\omega)$ that differ in the number of magnetic excitations in the intermediate state (either -1 or 1). After performing the integrations in Eq. (\ref{eq:Gsec_order}) as well as a transformation to momentum space we obtain to second order in ${\hat H}_{T}$
\begin{multline}
G(\omega) = \frac{J^2_{T}}{4}\sum_{k,q}\bigg[\frac{\la d_k d^\dagger_k\ra_0 \la r_q^\dagger r_q \ra_0}{(\varepsilon_k-\varepsilon_q)^2(\varepsilon_k-\varepsilon_q- \omega)} \\
+  \frac{\la d^\dagger_k d_k\ra_0 \la r_q r^\dagger_q \ra_0}{(\varepsilon_q-\varepsilon_k)^2(\varepsilon_q-\varepsilon_k- \omega)}\bigg],
\label{eq:G_sec_order2}
\end{multline}
which is valid both for AF and FM systems. Eq. (\ref{eq:G_sec_order2}) with $\omega = 0$ gives us $C_M$. When supplied with the relevant expectation values below, Eq. (\ref{eq:G_sec_order2}) tells us that the imaginary part of $G_\textrm{ret}(\omega)$ at small $\omega \ll E_D^+, E_D^-$ [which determines $R_M$, see Eq. (\ref{eq:response})] is zero to second order in ${\hat H}_{T}$. Hence, we need to analyze higher-order contributions to determine $R_M$. We will determine these contributions before analysing Eq. (\ref{eq:G_sec_order2}) in more detail.

We have explicitly checked that the only time ordering in the fourth-order expression for $G(\omega)$ that leads to an imaginary contribution at small $\omega \ll E_D^+, E_D^-$ is given by $\la \hat{H}'_{T}(t_1) \hat{N}'_D(t) \hat{H}'_{T}(t_2) \hat{H}'_{T}(t_3)\hat{N}'_D(0)\hat{H}'_{T}(t_4)\ra_0$. This leads to six unique terms that cannot be excluded a priori and differ in the number of magnetic excitations in the intermediate states. We illustrate the procedure followed to determine $G(\omega)$ by focussing on the term for which the excess number of magnetic excitations on the dot varies as $0 \to 1 \to 0 \to 1 \to 0$. After performing the integrations over time as well as a transformation to momentum-space we find the following contribution to $\textrm{Im}\left[G(\omega)\right]$ at small $\omega$ to fourth order in ${\hat H}_T$ due to this term:
\begin{equation}
\left(\frac{J_{T}}{2}\right)^4\sum_{k,\bar{k},q,\bar{q}}\frac{\la d^\dagger_k d_k d^\dagger_{\bar{k}} d_{\bar{k}}\ra_0\la r_q r^\dagger_{\bar{q}} r_{\bar{q}} r^\dagger_q\ra_0}{(\varepsilon_q - \varepsilon_k)^2(\varepsilon_q - \varepsilon_{\bar{k}})^2} \pi \delta(\varepsilon_q - \varepsilon_{\bar{q}} + \omega),
\label{eq:G_frth_order}
\end{equation}
where it is understood that we need to take the continuum limit on the reservoir in order for the delta-function to be well-defined. The other terms that make up $G(\omega)$ can be calculated analogously, and we will refrain from repeating the required steps here.

Up to this point, our results for $G(\omega)$ are identical for FM and AF systems. The sole difference between the two arises now from  the fact that $\la d_k^\dagger d_{\bar{k}} \ra_0$ and $\la r_k^\dagger r_{\bar{k}} \ra_0$ are different for AF and FM systems.

{\it AF case}. Assuming $T \to 0$, we put  $ \la r_k^\dagger r_{\bar{k}} \ra_0 = \la d_k^\dagger d_{\bar{k}} \ra_0 = \delta_{k,\bar{k}}\theta(\varepsilon_k - \mu_M)$. We can perform the summation over $q$ in Eqs. (\ref{eq:G_sec_order2})-(\ref{eq:G_frth_order}) by replacing $\sum_q \to \nu_R \int \ud q$, where $\nu_{R} = a | \partial \varepsilon_{q,R} / \partial q|^{-1}$ is the density of states in the reservoir. This leads to
\begin{equation}
C_M =t_\textrm{sd} \sum_{k} \frac{1}{(\varepsilon_k-\mu_M)^2} \textrm{ and } R_M = \frac{h}{2(g\mu_B)^2},
\label{eq:RCAF1}
\end{equation}
where $t_{\textrm{sd}} = (g\mu_B)^2 \nu_R (J_{T}/2)^2$ is the `magnetic transparency' of a small magnetic dot. We note that $C_M$ is well-defined since $|\varepsilon_k-\mu_M| \geq \textrm{min}[E_D^+,E_D^-]$.
For large dots, we can also perform the summation over $k$ using the density of states in the dot, $\nu_D$ (keeping in mind the previously discussed difficulties in the experimental realization of a finite $E_D^\pm$ for such dots). This leads to
\begin{equation}
C_M = t_\textrm{ld} \left[\frac{1}{E_D^+} - \frac{1}{E_D^-}\right] \textrm{ and } R_M = \frac{h}{(g\mu_B)^2},
\label{eq:RCAF2}
\end{equation}
where $t_\textrm{ld} =  (g\mu_B)^2 \nu_D \nu_R (J_{T}/2)^2$. In both cases we recover the fact that the spin resistance is universal in the sense that it does not depend on any microscopic parameters of the dot; not even on the coupling between the loop and the chain. As we show in the appendix, this result is related to the fact that one can map the AF spin-1/2 chain to one-dimensional fermions with interactions. This model has been studied extensively in the past few years in the context of electronic RC circuits\cite{Buttiker1993,Gabelli2006,Nigg2006,Nigg2008,Mora2010,Hamamoto2010,Filippone2011,Filippone2012,Gabelli2012} even beyond perturbation theory  using for instance bosonization or Monte-Carlo calculations. Therefore, this mapping allows us to extend our results to any value of the tunneling coupling.
Furthermore, the effect of electron-electron interactions in the reservoir and dot on the parameters of the RC circuit have been analyzed using a Luttinger Liquid description\cite{Mora2010,Hamamoto2010}. Since these interactions translate to a $S_i^zS_{i+1}^z$-term in the spin chains, these results allow us to extend Eqs. (\ref{eq:RCAF1})-(\ref{eq:RCAF2}) to finite values of $\Delta_{D(R)} \leq 1$. A finite value of $\Delta_{D(R)}$ corresponds to a deviation of the Luttinger Liquid parameter from the noninteracting value $K=1$. This simply leads to an additional factor $1/K$ in the result for $R_M$, so the resistance remains universal. The value of the capacitance $C_M$ is changed in a nontrivial manner\cite{Mora2010,Hamamoto2010}.

{\it FM case}. The zero temperature limit is pathological for FM systems, since the Bose-Einstein distribution diverges at $T=0$. Therefore, we will consider finite (but small) temperatures. Specifically, we will assume that $k_B T \ll E_D^+$, so that we can put $\la d_k^\dagger d_{\bar{k}} \ra_0 = 0$. Furthermore, we linearize $\la r_q^\dagger r_{\bar{q}} \ra_0 = n_B(E_R^+ + \epsilon_q$) around the minimal value of the energy spectrum $E_R^+$. Substitution of these values into Eqs. (\ref{eq:G_sec_order2})-(\ref{eq:G_frth_order}) yields 
\begin{equation}\begin{split}
C_M & = \frac{t_\textrm{ld}}{2} \frac{n_B(E_R^+)}{E_D^+-E_R^+}\frac{\delta}{E_D^+-E_R^+}\,,\\
R_M & = -\frac{2h}{(g\mu_B)^2} \left(\frac{E_D^+-E_R^+}{\delta}\right)^2\,,
\end{split}
\end{equation}
where $\delta = | n_B(E_R^+) / n'_B(E_R^+)|$. We find then that in the bosonic case the spin resistance is no longer universal, although it is still independent of the tunneling coupling $J_T$ (at least to fourth order in perturbation theory). The fact that the relaxation resistance is negative is not a fundamental issue but only means that the dynamical response of the spin capacitor is out of phase with the perturbation.

Note that the universality of the relaxation resistance in electronic interacting systems was shown in Ref. \onlinecite{Mora2010} to be intimately related to the Korringa-Shiba relation\cite{Shiba1975} (see also Garst et al.\cite{Garst2005} for an extended version of this relation) which relates the imaginary part of the charge susceptibility to the square of its real part. Such a relation applies at or near a Fermi-liquid fixed point. It is therefore not surprising to find a non-universal behavior for  $R_M$ in the FM case, where low-energy excitations are bosonic.
\\

We turn now to the possibility of using the setup displayed in Fig. 1 as a source of magnetic quasiparticles. F\`eve {\it et al.} have shown~\cite{Feve2007} experimentally that the electronic quantum RC circuit can be used as an on-demand single-electron source, by applying a square voltage pulse $V_D(t)$ with a large amplitude to the dot, such that the system is taken beyond the linear response regime. The Jordan-Wigner transformation shows that the AF system is equivalent to the electronic system, and hence we propose that the AF system can be used as an on-demand single-spinon source, with the simple substitution  $eV_D(t) \to g\mu_BB_D(t)$. For FM systems, the situation is fundamentally different; due to the fact that the (bosonic) thermal distribution of magnons allows for more than one magnon per momentum state, it is not possible to create a single-magnon source using the same mechanism. However, an on-demand few-magnon source appears feasible.

Finally, we comment on the possibility of measuring the properties of the magnetic quantum RC circuit experimentally. The ultimate implementation uses parallel spin chains, such as depicted in Fig. 1(a). Furthermore, molecular magnets\cite{Gatteschi2007,Sessoli1993,Thomas1996,Friedman1996,Leuenberger2001,Ardavan2007} could be a good candidate to take the role of magnetic dot due to their beneficial properties, such as their increased size, chemical engineerability, and the possibility to control the spin state using electric- instead of magnetic fields\cite{Trif2008}. However, based on the magnitude of the spin currents and the involved time scales, the implementation of the magnetic RC circuit in the above systems appears challenging. Therefore, we propose an alternative system to test our predictions, namely ultracold atoms in optical lattices.

It has been shown\cite{Kuklov2003,Duan2003} that ultracold atoms trapped in optical lattices can be used to implement effective spin-1/2 Heisenberg chains (both AF and FM) with tunable anisotropy $\Delta_{R(D)}$. Furthermore, it is now possible\cite{Bakr2010,Sherson2010,Fukuhara2013} to measure the spin state in such systems dynamically, locally, and with single-spin precision.

We assume that the (effective) magnetic field $B_D(t)$ has the form $B_D(t) = B_D\cos (\omega_0 t)$. Validity of our results then requires $B_D,\hbar\omega_0 \ll |E_D^{\pm}|$. From Eq. (\ref{eq:response}) it follows that the resulting magnetization $M_D(t)= M_D^0 \cos(\omega_0 t) - M_D^1 \sin(\omega_0 t)$, where $M_D^0  = C_M B_D$ and $M_D^1 = R_MC_M^2 \omega_0 B_D$. To measure $R_M, C_M$, one has to be able to distinguish between these two contributions. For simplicity, we will give numbers for AF systems and small dots. Using Eq. (\ref{eq:RCAF1}) and taking the continuum limit on the dot, we estimate $M_D^0 \sim \xi (g\mu_B)$ and $M_D^1 \sim \xi^2 (\hbar \omega_0 /g\mu_B B_D) (g\mu_B)$, where $\xi = (J_{T}/J_D)(J_{T}/J_R)(g\mu_B B_D/E_D^+)$ is a small fraction. We have assumed that $|E_D^+| = |E_D^-|$ and $\varepsilon_{k,D(R)} = -J_{D(R)} \cos (k a)$. If we assume $J_{T} \lesssim J_D \approx J_R$ (see the discussion below Eq. (\ref{eq:RCAF2}) for the validity of this approximation) and $\hbar \omega_0 \approx g \mu_B B_D = 0.1 E_D^+$, it follows that a collection of $\sim 10^2$ parallel chains suffices to determine $R_M, C_M$ in repeated measurements. For the smallest magnetic dots, $E_D^+ \approx J_D$. A representative value\cite{Duan2003} for $J_D$ is $J_D/\hbar \sim 0.1$ kHz, which leads to $\omega_0 = 10$ Hz, smaller than the typical lifetime of excitations in such systems\cite{Fukuhara2013}.

{\it Conclusion}. In conclusion, we have studied a model of a magnetic RC circuit and computed the effective capacitance and resistance from microscopic parameters. We have shown that the spin resistance is universal for AF coupling but not for FM systems. Furthermore, we have shown that our predictions can be presently tested with time-resolved experiments in ultracold atoms in optical lattices. This opens the path towards the realization of on demand single spin excitation (spinon or magnon) emitters that would be one of the key ingredients for spintronics.\\

{\it Acknowledgements}. K. v. H. and D. L. acknowledge support from the Swiss NSF, the NCCR QSIT, and the FP7-ICT project "ELFOS". M. A. and P. S. acknowledge support from the RTRA project NanoQuFluc and the ERC starting Independent Researcher Grant NANOGRAPHENE 256965.

\appendix

\section{Out-of-plane terms in the hopping Hamiltonian}
In this section we will discuss the effect of a term $H_Z = J_Ts_N^z S_0^z$ on $G_\textrm{ret}(\omega)$. Such a term describes the interaction between the out-of-plane components of the boundary spins ${\bf s}_N$ and ${\bf S}_0$, and corresponds to a local four-body interaction in the language of the magnetic quasiparticles. The main question is whether such interaction terms give a contribution to the imaginary part of $G_\textrm{ret}(\omega)$ at small $\omega$, possibly even in an order that is lower than the fourth order contribution we found in the main text. We will show explicitly that this is not the case.

When taking both $H_T$ and $H_Z$ into account, we should simply replace the propagator in Eq. (6) of the main text by $\hat{U}(-\infty,\infty) = T_{t_1}\textrm{exp}\{-i\int_{-\infty}^\infty \ud t_1 [\hat{H}_T'(t_1)+\hat{H}_Z'(t_1)]\}$ and perform the expansion of Eq. (6) in the main text in powers of $J_T$ using this new propagator.

Since $H_Z$ contains an equal number of creation and annihilation operators (both on the dot and in the reservoir), we need to consider also the odd orders in the expansion of Eq. (6) in the main text in powers of $J_T$. The first order contribution to $G(\omega)$ due to $H_Z$ is given by
\begin{equation}
G(\omega) = -\int_0^\infty  \ud t e^{i\omega t} \int_{-\infty}^\infty \ud t_1 \la T_t\hat{H}'_Z(t_1)N'_D(t)N'_D(0)\ra_0.
\end{equation}
It can immediately be seen that the contribution to $G_\textrm{ret}(\omega)$ due to this term evaluates to zero. This is due to the fact that the $S_0^z$ operator does not change the number of magnetic quasiparticles $N_D$ on the dot, and we defined $N_D$ such that $N'_D(t)|\textrm{gs}\ra = 0$. From this reasoning it follows that all contributions that contain only $H_Z$-operators vanish, regardless  of the order.

The first nonzero contribution to $G(\omega)$ that contains both $H_Z$ and $H_T$ operators (and hence cannot be said to be zero a priori) is third order in $J_T$. The contribution to the third-order expansion originating from the numerator in Eq. (6) of the main text is given by
\begin{multline}
G(\omega) = -\int_0^\infty  \ud t e^{i\omega t} \int_{-\infty}^\infty \ud t_1\int_{-\infty}^\infty \ud t_2\int_{-\infty}^\infty \ud t_3 
\\ \times \la T_t\hat{H}'_Z(t_1)\hat{H}'_T(t_2)\hat{H}'_T(t_3)N'_D(t)N'_D(0)\ra_0.
\label{eq:3rd_order}
\end{multline}
Clearly, the expansion of the denominator of Eq. (6) in the main text also yields third-order contributions. However, those only serve to cancel the disconnected terms in Eq. (\ref{eq:3rd_order}). To calculate the third order contribution to $G(\omega)$, we proceed as in the main text; we explicitly write out the different time orderings, and perform the integrations over time as well as a transformation to momentum space for each term individually. We find no contribution to $\textrm{Im}[G_\textrm{ret}(\omega)]$ at small $\omega$. We do find (real-valued) contributions to $G_\textrm{ret}(\omega)$ at $\omega=0$. However, since the largest contribution to $G_\textrm{ret}(0)$ is only second order in $J_T$, these contributions simply renormalize the capacitance $C_M$ of the circuit.

Lastly, we can calculate the fourth order contributions to $G(\omega)$ that contains $H_Z$ as well as $H_T$
\begin{multline}
G(\omega) = i\int_0^\infty  \ud t e^{i\omega t} \int_{-\infty}^\infty \ud t_1 ... \int_{-\infty}^\infty \ud t_4 \\
\times  \la T_t\hat{H}'_Z(t_1)\hat{H}'_Z(t_2)\hat{H}'_T(t_3)\hat{H}'_T(t_4)N'_D(t)N'_D(0)\ra_0.
\end{multline}
Similar to the third order contributions, we find no contribution to $\textrm{Im}[G_\textrm{ret}(\omega)]$ at small $\omega$; we do again find contributions to $G_\textrm{ret}(0)$ which renormalize $C_M$.

\section{Antiferromagnetic case}
 
One can map the AF spin-$1/2$ Heisenberg spin chain onto a system of interacting one-dimensional fermions using the Jordan-Wigner transformation\cite{Giamarchi2003}. This transformation is defined by $S_i^+=c_i^\dagger e^{i\pi \sum_{j=-\infty}^{i-1}c_j^\dagger c_j}$ and $S_i^z=c_i^\dagger c_i-1/2$, where $S^{\pm}=S_x\pm iS_y$ and $c_i$ is the fermionic annihilation operator. With this transformation, Eq. (2) in the main text can be rewritten as
  \begin{equation}
    \begin{split}
      H& = \frac{J_D}{2}\sum_i[c_{i+1}^\dagger c_i+\textrm{h.c.}]+g\mu_B [B_D^0+B_D(t)]\sum_i (c_i^\dagger c_i-1/2)\\
      & + \sum_i J_D\Delta_D (c_{i+1}^\dagger c_{i+1}-1/2)(c_i^\dagger c_i-1/2)\,.
    \end{split}
  \end{equation}
If we now consider a dot coupled to a semi-infinite chain we can apply the same transformation to both parts of the system. In this language the system is equivalent to the electronic quantum capacitor (with $(g\mu_B)^2 \leftrightarrow e^2$ and $g\mu_B B \leftrightarrow eV $) that has been studied extensively in the literature\cite{Buttiker1993,Nigg2006,Nigg2008,Gabelli2006,Mora2010,Hamamoto2010,Filippone2011,Filippone2012}. In particular, it has been\cite{Buttiker1993} shown and observed experimentally\cite{Gabelli2006} that the charge relaxation resistance without interactions (\textit{i.e.} with $\Delta_{D(R)}=0$ is universal at zero temperature, as long as the spectrum in the dot remains discrete. When interactions are turned on in the dot this result remains valid for small dots (with a discrete spectrum) and leads to another universality class for large dots (with a continuous spectrum) thanks to the Coulomb gap\cite{Mora2010}. Here, universal means that the value of the charge relaxation resistance does not depend on any microscopic parameters, such as the coupling between the dot and the reservoir. Instead, it is simply a combination of fundamental constants. In the AF language the charge relaxation resistance becomes $R_M=h/2(g\mu_B)^2$ for a small dot and $R_M=h/(g\mu_B)^2$ for a large dot. In the main text we derive these results in the weak-coupling limit, but the close resemblance between the AF- and electronic systems allows us to extend them to the full range of parameters\cite{Mora2010}.

When interactions are also present in the reservoir, Hamamato \textit{et al.}\cite{Hamamoto2010} have shown, using the Luttinger liquid framework\cite{Giamarchi2003}, that the charge relaxation resistance is re-scaled by a factor $1/K$, where $K$ is the Luttinger interaction parameter. $K=1$ for non-interacting particles and $K<1$ for repulsive interactions. However, this result only holds for weak-enough interactions, depending of the strength of the tunneling coupling. A Kosterlitz-Thouless phase transition separates a coherent phase (in which the charge relaxation resistance is universal) to an incoherent phase (in which $R_M$ is no longer quantized). For a fully open channel (infinite coupling between dot and reservoir) the critical Luttinger parameter is $K_c=1/2$. Interestingly, in the magnetic language, this corresponds to the opening of a gap in the spectrum of the infinite chain as the system goes from the conducting $XY$ phase to the Mott-insulating Ising phase. Indeed, the Bethe-Ansatz solution gives\cite{Giamarchi2003}

\begin{equation}
  \Delta=-\cos\pi\beta^2\,,\quad \frac{1}{K}=2\beta^2\,.
\end{equation}
Therefore $K=1/2$ corresponds to $\Delta=1$, which is the boundary between the two phases mentioned above.

\section{Ferromagnetic case}

Ferromagnetic systems with spin $S \gg 1$ are handled using the so-called Holstein-Primakoff transformation\cite{Mattis1981}. This transformation allows us to describe the spin excitations as magnons: magnetic quasiparticles that are bosonic in nature. The bosonic operators $d_j, d_j^\dagger$ in the dot are implicitly defined as
\begin{equation}\label{HP_trans}
    \begin{split}
      S_j^z =&  S-d^\dagger_j d_j,\\
      S_j^+=& \sqrt{2S}\sqrt{1-d^\dagger_j d_j/2S} \,d_j, \\
      S_j^-=& \sqrt{2S}\,d^\dagger_j\,\sqrt{1-d^\dagger_j d_j/2S},\\
    \end{split}
  \end{equation}
and similar expressions hold for the operators in the reservoir. In the large-$S$ limit, we can expand the square roots in powers of $1/S$. If we take only the lowest-order terms into account, we can diagonalize the Hamiltonian in terms of free magnons. This transformation has to be applied to both parts of the system independently.

We found in the main text that the resistance $R_M$ of our circuit is determined by the fourth order (in $J_T$) contribution to $G_\textrm{ret}(\omega)$. However, the nonlinear terms in the expressions for $S_j^{\pm}$ yield contributions to $G_\textrm{ret}(\omega)$ that superficially appear similar to the fourth order contributions.  Specifically, they contain the same amount of bosonic creation-/annihilation operators. Therefore, we have checked explicitly that the non-linearities in (\ref{HP_trans}), when treated perturbatively, do not affect the physics of our magnetic quantum RC circuit qualitatively. In particular, they do not cause dissipation (imaginary part of the response function) to second order. Instead, they only renormalize the correlators. To be more precise, we found that the sole effect of substituting $S_0^+ \approx \sqrt{2S}[1-d^\dagger_0 d_0/4S]d_0$ (and the accompanying substitution for $S_0^-$) in the expression for $G(\omega)$ given by Eq. (7) in the main text is to renormalize the final result for $C_M$ with a factor $1-\la d^\dagger_0 d_0\ra_0/2S$. Similarly, when replacing $s_N^+ \approx \sqrt{2S}[1-r^\dagger_N r_N/4S]r_N$, we find that the capacitance $C_M$ is renormalized by a factor $1-\la r^\dagger_N r_N\ra_0/2S$.

\end{document}